\documentclass{article}

    \PassOptionsToPackage{numbers, compress}{natbib}
    \usepackage[final]{neurips_2025}

\usepackage{caption}  
\usepackage{makecell}

\usepackage[utf8]{inputenc} % allow utf-8 input
\usepackage[T1]{fontenc}    % use 8-bit T1 fonts
\usepackage{hyperref}       % hyperlinks
\usepackage{url}            % simple URL typesetting
\usepackage{booktabs}       % professional-quality tables
\usepackage{amsfonts}       % blackboard math symbols
\usepackage{nicefrac}       % compact symbols for 1/2, etc.
\usepackage{microtype}      % microtypography
\usepackage{xcolor}         % colors
\usepackage{booktabs}
\usepackage{threeparttable}
\usepackage{graphicx}
\usepackage{multirow}
\usepackage{algorithm}
\usepackage{algorithmic}
\usepackage{amssymb}
\usepackage{pifont}

\title{CoVoMix2: Advancing Zero-Shot Dialogue Generation with Fully Non-Autoregressive Flow Matching}

\author{%
  \small{Leying Zhang}$^{1,2}$\thanks{Work done during an internship at Microsoft Azure AI. zhangleying@sjtu.edu.cn} 
  \And \small{Yao Qian}$^{2}$\thanks{Correspondence: yaoqian@microsoft.com} 
  \And \small{Xiaofei Wang}$^{2}$ 
  \And \small{Manthan Thakker}$^{2}$
  \And \small{Dongmei Wang}$^{2}$
  \And \small{Jianwei Yu}$^{2}$
  \And \small{Haibin Wu}$^{2}$
  \And \small{Yuxuan Hu}$^{2}$
  \And \small{Jinyu Li}$^{2}$
  \And \small{Yanmin Qian}$^{1}$
  \And \small{Sheng Zhao}$^{2}$
  \And
  $^{1}$\texttt{Shanghai Jiao Tong University,  China} 
  \And
  $^{2}$\texttt{Microsoft, USA} 
}

\begin{document}

\maketitle

\begin{abstract}
Generating natural-sounding, multi-speaker dialogue is crucial for applications such as podcast creation, virtual agents, and multimedia content generation. However, existing systems struggle to maintain speaker consistency, model overlapping speech, and synthesize coherent conversations efficiently. In this paper, we introduce CoVoMix2, a fully non-autoregressive framework for zero-shot multi-talker dialogue generation. CoVoMix2 directly predicts mel-spectrograms from multi-stream transcriptions using a flow-matching-based generative model, eliminating the reliance on intermediate token representations. To better capture realistic conversational dynamics,  we propose transcription-level speaker disentanglement, sentence-level alignment, and prompt-level random masking strategies.  Our approach achieves state-of-the-art performance, outperforming strong baselines like MoonCast and Sesame in speech quality, speaker consistency, and inference speed. Notably, CoVoMix2 operates without requiring transcriptions for the prompt and supports controllable dialogue generation, including overlapping speech and precise timing control, demonstrating strong generalizability to real-world speech generation scenarios. Audio samples are available~\footnote{\url{https://www.microsoft.com/en-us/research/project/covomix/covomix2/}}. 
\end{abstract}

\begin{figure}[htbp]
  \centering
\centerline{\includegraphics[width=1.0\linewidth]{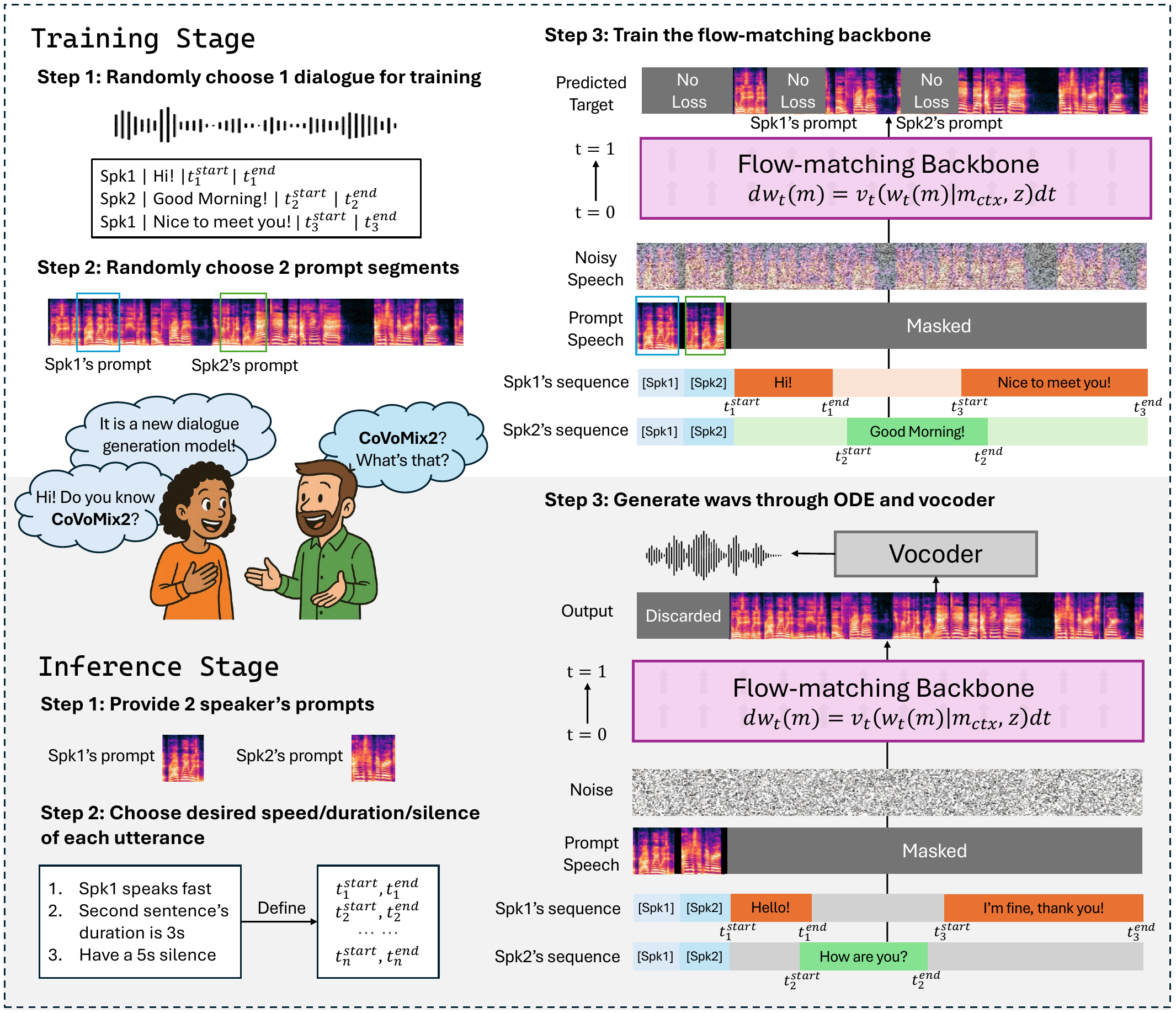}}
\caption{The overview of the proposed CoVoMix2 framework
}
\label{fig:intro}
\end{figure}

\section{Introduction}

The field of speech synthesis has witnessed remarkable progress in recent years, particularly in zero-shot text-to-speech (TTS) systems that can generate high-quality, natural-sounding speech in voices not seen during training~\cite{ju2024naturalspeech, leng2023prompttts,lajszczak2024base,du2024cosyvoice,anastassiou2024seed,valle,lee2024voiceldm,lee2024ditto}. These advancements have enabled applications such as personalized virtual assistants, audiobook narration, and interactive voice response systems. However, extending these capabilities to multi-talker dialogue generation, where multiple speakers engage in natural, dynamic, multi-turn conversations, remains a significant challenge. 

Conventional approach to synthesize a dialogue is by generating multiple monologues sequentially and concatenating them, which often results in unnatural interactions and poor speaker coordinations~\cite{nguyen2023generative, sesameCrossingUncanny,zhang2024covomix, ju2025MoonCast}. Recent efforts have shifted towards generating the entire dialogues in a more integrated manner. CoVoMix~\cite{zhang2024covomix} was the first attempt to employ a multi-stream autoregressive (AR) text-to-semantic model and a non-autoregressive (NAR) acoustic model to synthesize mixed mel-spectrograms for zero-shot dialogue generation. 
NotebookLM~\cite{deepmindNotebookLM,borsos2023soundstorm} leverages hierarchical transformer to produce a stream of audio tokens autoregressively, which are decoded back to a dialogue waveform. MoonCast~\cite{ju2025MoonCast} utilizes long-context text-to-semantic model to support coherent dialogue generation. These systems typically adopt a two-stage AR+NAR pipeline involving intermediate representations, such as semantic or audio tokens.

Despite these advances, current systems face several key limitations. Firstly, dependence on AR components and intermediate representations introduces considerable complexity and slow inference speed. Secondly, the lack of fine-grained control over prosodic features, such as speaking rate, utterance duration, overlap, and pauses, often results in unnatural and less coherent dialogue flow. Thirdly, speaker identity inconsistencies, where the wrong voice is occasionally assigned to an utterance, undermine perceived authenticity. Lastly, current methods~\cite{nguyen2023generative,zhang2024covomix,lu2025slide, mitsui2023towards} are hindered by their reliance on stereo audio for training and paired audio-text prompts for inference, often necessitating external ASR models and limiting scalability.

To address these challenges, we propose CoVoMix2 (\textbf{Co}nversational \textbf{Vo}ice \textbf{Mix}ture Generation), a novel framework for zero-shot multi-talker dialogue generation based on a fully NAR flow-matching approach.  Our main contributions are summarized as follows:

\begin{enumerate}
  
 \item We exploit a fully non-autoregressive framework for zero-shot multi-talker dialogue generation. It is an end-to-end modeling from the interleaved character sequence of two speakers to their mixed mel-spectrograms, with intrinsic alignment between characters and mel-spectrogram frames.
 \item We propose a transcription-level disentanglement strategy in which each speaker’s transcriptions are provided as separate streams, enabling potential control over the timing of generated overlapping speech.
 \item Additionally, we introduce sentence-level alignment and prompt-level masking strategies that facilitate accurate speaker identity assignment and eliminate the need for intermediate representations or transcriptions of speaker prompts during inference.
 \item Extensive experiments show that CoVoMix2 outperforms strong baselines, achieving state-of-the-art (SOTA) performance among open-source checkpoints, while requiring significantly less training data and delivering faster inference speed.
 
\end{enumerate}

\section{Related Work}

\subsection{Flow-matching based Speech Synthesis}
Flow-matching-based speech synthesis has emerged as a promising approach in the field of text-to-speech (TTS) systems~\cite{chen2018cnf,lipman2023flow}. It models speech generation as a continuous transformation from simple distributions to complex speech representations, enabling fast, parallel inference and improved controllability compared to traditional AR models~\cite{valle,han2024vall}.

Several recent models~\cite{mehta2024matcha,le2023voicebox,vyas2023audiobox,leyingBPBR,guo2024voiceflow,10832278} have demonstrated the effectiveness of flow matching across tasks such as zero-shot TTS, speech inpainting, and voice conversion. Notably, E2-TTS~\cite{eskimez2024e2} and F5-TTS~\cite{chen2024f5} remove the need for phoneme alignment or explicit duration modeling, resulting in simpler pipelines with high-quality output. These advancements establish flow-matching as a promising direction for building scalable and efficient TTS systems.

\subsection{Multi-talker Dialogue Generation}
Multi-talker dialogue generation aims to synthesize entire multi-turn conversations between multiple speakers, where each utterance must be coherent, contextually relevant, and speaker-specific. This task is significantly more complex than single-speaker synthesis due to the need for speaker turn-taking, identity preservation, and interaction dynamics.

Early approaches, such as CoVoMix~\cite{zhang2024covomix}, introduced a multi-stream AR text-to-semantic model combined with a NAR acoustic decoder to generate dialogues in a zero-shot manner. Other systems like Soundstorm~\cite{borsos2023soundstorm} and NotebookLM~\cite{deepmindNotebookLM} incorporate both AR and NAR components to model hierarchical audio tokens or contextual embeddings. 
Further, works such as Chats\cite{mitsui2023towards} and SLIDE\cite{lu2025slide}, built upon the dGSLM framework~\cite{nguyen2023generative}, leverage dual-tower transformer architectures to capture interleaved speaker information. Encoder-decoder models like Dia 1.6B\cite{githubGitHubNarilabsdia} and Parakeet\cite{darefsky2024parakeet} directly predict audio codec tokens, which are then decoded into waveforms. MoonCast~\cite{ju2025MoonCast} addresses long-context dialogue generation using a transformer-based language model, concatenating acoustic prompts at each turn to manage speaker changes.

While these methods have advanced the field by improving naturalness and speaker consistency, most rely on AR decoding or hybrid AR/NAR architectures. These designs introduce several drawbacks: slow inference due to step-by-step generation, reliance on complex intermediate representations, limited controllability over conversational dynamics such as overlap, pauses, or timing, and dependence on the paired text-audio speaker prompts.

\subsection{Conversational Speech Synthesis}
Conversational Speech Synthesis (CSS) focuses on generating individual utterances that are contextually appropriate within a dialogue, typically from a single speaker. Unlike full dialogue generation, CSS produces each utterance one-by-one, conditioned on previous dialogue history, rather than synthesizing an entire conversation simultaneously~\cite{hu2024fctalker, xue2023m,guo2021conversational,lee2023dailytalk}.

For instance, GPTTalker~\cite{liu2024generative} models multimodal dialogue context by converting history into discrete tokens processed by a language model to generate expressive, context-aware responses.
Sesame~\cite{sesameCrossingUncanny} adopts two AR transformers~\cite{defossez2024moshi}: a multimodal transformer backbone processes text and audio tokens, followed by a audio decoder transformer that reconstructs high-quality speech. Although Sesame supports dialogue-style synthesis, it fundamentally generates each speaker’s utterance sequentially given the previous dialogue context, rather than modeling simultaneous multi-speaker interaction.

While CSS methods achieve high expressiveness and coherence for individual utterances, they fall short in handling the broader structure and dynamics of real-time, overlapping multi-speaker dialogue.

Our proposed CoVoMix2 differs fundamentally from both traditional multi-talker dialogue generation systems and CSS approaches. Unlike prior work that relies on AR or hybrid pipelines, or CSS systems that synthesize one speaker’s utterance at a time, CoVoMix2 enables fully NAR, simultaneous generation of multi-speaker dialogues. It directly predicts mel-spectrograms from disentangled multi-stream transcriptions, allowing efficient, accurate, and controllable synthesis with better support for real conversational dynamics such as overlapping speech and natural pauses.

\section{CoVoMix2}

Zero-shot dialogue generation aims to synthesize multi-speaker conversations in voices not encountered during training. We propose CoVoMix2, a fully NAR zero-shot dialogue generation model that directly generates mel-spectrograms from raw dialogue transcriptions. As shown in Figure~\ref{fig:intro}, CoVoMix2 operates without intermediate representations such as phonemes or audio tokens, enabling efficient and scalable zero-shot dialogue generation.

Let the training dataset be denoted as $D = \{x, y\}$, where $x$ is a dialogue waveform containing utterances from two speakers, and $y=[y_1,...,y_n]$ is the corresponding transcription.  Each transcription segment $y_i$ is annotated with the content $T_i$, speaker label $s_i$,  the start and end time $t_i^{start}, t_i^{end}$. The corresponding mel-spectrogram of $x$ is noted as $m$. To support simultaneous, zero-shot, multi-speaker dialogue generation, we introduce three key design strategies in the following sections, yielding a pair of input text sequences $z =  [z_1,z_2]$, each representing a separate speaker stream, and the  acoustic prompts $m_{ctx}$ of the target speakers.

\subsection{Transcription-Level Speaker Disentanglement}

Natural speaker switching and overlapping speech are essential features in realistic dialogue generation. Prior work often inserts speaker-change tokens (e.g., [spkchange]) to indicate a switch between speakers~\cite{zhang2024covomix, ju2025MoonCast,deepmindNotebookLM,borsos2023soundstorm}. However, single-stream representations entangle multiple speakers’ utterances into a flat sequence, making it difficult for the model to explicitly capture speaker-specific timing, and overlap. Moreover, this design creates ambiguity in conditioning, particularly when aligning acoustic prompts to textual content. In a shared text stream, the model must disentangle which portions belong to which speaker and match them to the correct prompt, increasing the risk of identity confusion.

\begin{figure}[htbp]
  \centering
\centerline{\includegraphics[width=1.0\linewidth]{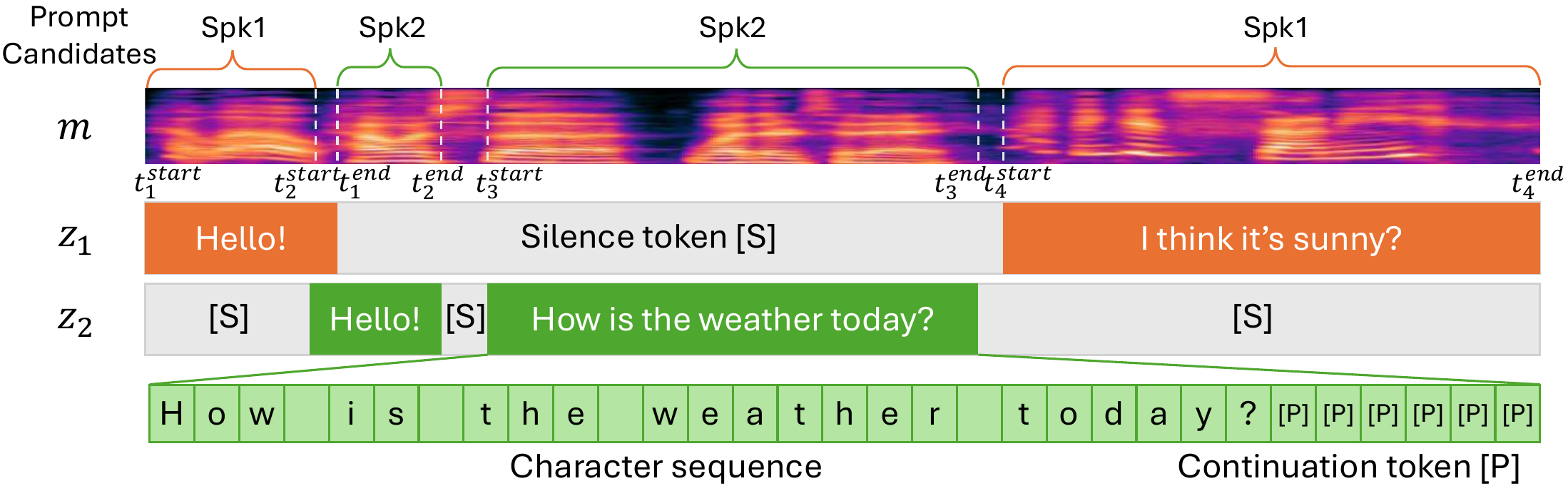}}
\caption{Example of the input data organization}
\label{fig:method}
\end{figure}

Instead, as shown in Figure~\ref{fig:method}, we propose a more structured approach by disentangling the transcript into multiple parallel streams, one per speaker. Each text stream $z_i$ contains two types of content: Active speech segments, defined by their respective time intervals $[t_i^{start}, t_i^{end}]$. and Silence intervals, denoted using a special token $[S]$.  These streams enable precise temporal control over individual speaker utterances, including overlapping and silence. This design grants fine-grained control over the interaction structure, improving naturalness and flexibility.

\subsection{Sentence-Level Alignment}
High-quality alignment at the phoneme level is often unavailable in real-world data~\cite{zhang2024covomix}. Likewise, modeling intermediate representations (e.g., codec tokens) is resource-intensive and requires large-scale training data. As an alternative, inspired by recent models~\cite{eskimez2024e2, chen2024f5}, we design the sentence-level alignment that is suitable for both monologue and dialogue scenarios. 

As indicated in Figure \ref{fig:method}, each active speech segment in the speaker stream is converted into a sequence of characters (including upper and lower cases letters and punctuation). For the rest of the active speech segment, we add a special continuation token $[P]$ for padding. These sequences, temporally anchored by the corresponding start and end times, serve as the text input for the mel-spectrogram prediction. The model learns the intrinsic alignment between characters and mel-spectrogram frames and synthesizes each utterance within its designated time range without explicit duration prediction.

\subsection{Prompt-Level Random Masking}

Accurate voice conditioning is critical to avoid speaker confusion. Previous approaches either maintained parallel conditioning streams~\cite{zhang2024covomix, lu2025slide}, used concatenated speaker prompt for continuation~\cite{borsos2023soundstorm} or appended speaker prompts at each dialogue turn to guide speech synthesis~\cite{ju2025MoonCast}. We introduce a prompt-level random masking strategy to ensure robust and diverse speaker conditioning.

As shown in Figure \ref{fig:intro} and \ref{fig:method}, for each training sample, we first find all the available monologue segments for each speaker as the prompt candidates. We then randomly select a prompt segment from the candidates. The chosen prompts are concatenated at the very beginning with the masked training sample using a separator token to construct the prompt sequence $m_{ctx}$. Moreover, in the text stream $z$, we use special tokens $[Spk1]$ and $[Spk2]$  as indicators (not text tokens) to replace the transcription of the prompt, distinguishing which segments belong to each speaker. 

Furthermore, to prevent prompt leakage, where the model copies the prompt directly into the output, we exclude the prompt region from the loss computation. This encourages the model to generalize voice characteristics rather than memorize the prompt audio.

\subsection{Flow-Matching-Based Mel Spectrogram Generation}

Our model employs Flow Matching (FM), a simulation-free training method for Continuous Normalizing Flows (CNFs)~\cite{lipman2023flow, chen2018neural}. It is a class of generative models that learn to transform a simple distribution (e.g., Gaussian noise) into a complex data distribution (e.g., mel-spectrograms) through a continuous mapping. This mapping is defined by solving an Ordinary Differential Equation (ODE).

Specifically, the model learns the distribution $q(m | z, m_{ctx})$ where $m$ is the target mel-spectrogram, $z$ is the speaker-aware text streams and $m_{ctx}$ is the acousic prompts. At training time, a noise sample $m_0$ is drawn from a standard Gaussian distribution. The training objective is to minimize the L2 distance between the predicted and true flow, given by Eq.\ref{eq:loss}, where $v_t(\cdot)$ is the model's predicted vector field at time $t\in[0,1]$,  $w=(1-(1-\sigma_{min})t)m_0 +tm$, and $\sigma_{min}$ is a hyper-parameter to control the deviation of flow-matching. Notably, the loss is not computed on segments that originate from the prompt region, which is excluded using a masking function 
 $M(\cdot)$.
\begin{equation}
    \mathcal{L} = \mathbb{E} \Vert M( (m - (1 - \sigma_{min})m_0) - v_t(w, m_{ctx}, z ; \theta)) \Vert^2
    \label{eq:loss}
\end{equation}
We also apply Classifier-Free Guidance (CFG)~\cite{ho2022classifier, le2023voicebox} to improve sample quality by interpolating between conditioned and unconditioned flows. During training, the acoustic prompt $m_{ctx}$ and text sequences $z$ are dropped with $p_{uncond}$. 

During inference, the CFG vector field becomes Eq.\ref{eq:cfg}, with $\alpha$ controlling the strength of guidance. Durations for each utterance are computed based on syllable counts and a predefined speaking rate, allowing the construction of the input text streams $z$. A mel-spectrogram is then generated by sampling noise $m_0$ and solving the ODE defined by the flow field.
\begin{equation}
    \tilde{v}_t(w, m_{ctx}, z ; \theta) = (1 + \alpha) v_t(w, m_{ctx}, z ; \theta) - \alpha \tilde{v}_t(w; \theta)
    \label{eq:cfg}
\end{equation}

\subsection{Training Strategy: Curriculum Learning and Data Mixing}

To enable the model with dialogue generation capability efficiently, we adopt a two-stage curriculum learning strategy during training. First, the model is pretrained on high-quality monologue datasets, which helps it learn accurate pronunciation and acoustic modeling. In contrast, directly training on multi-speaker data causes degraded output quality, including mispronunciations and unintelligible speech. Then, in the second stage, we train the model on multi-speaker dialogue datasets to enhance the dialogue generation capability. 

To improve robustness and generalization, built on the large scale of monologue dataset, we design the data mixing strategy, using various sources of data during the second stage training. Specifically, we mix data from several sources, including ASR-transcribed podcast dialogues, audiobook-style single-speaker datasets and simulated overlapped dialogues to support overlapping capability. Benefited from this data diversity, CoVoMix2 does not require any human-annotated dialogues to get satisfied and natural results. For further enhancement, we demonstrate in Appendix \ref{sec:finetune} that fine-tuning the model with just 20 minutes of clean, human-annotated dialogue can significantly boost performance.

\section{Experimental Setup}

\subsection{Training and Inference Data Preparation}

\begin{figure}[htbp]
  \centering
\centerline{\includegraphics[width=1.0\linewidth]{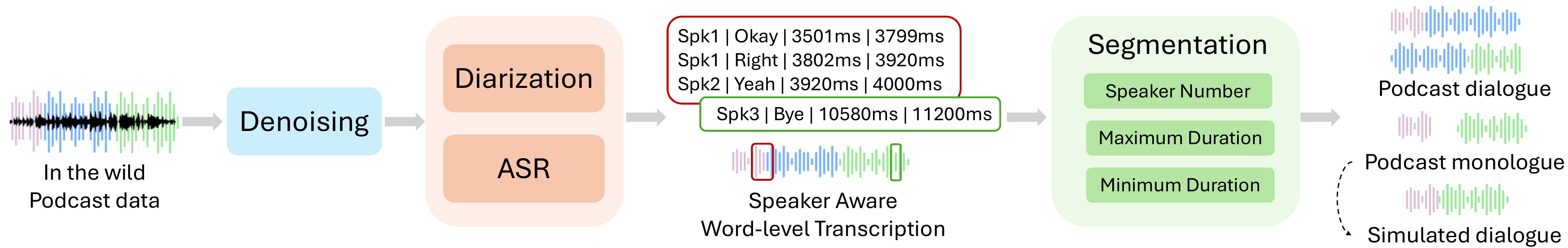}}
\caption{Data processing pipeline
}
\label{fig:dataprocessing}
\end{figure}

To train CoVoMix2, we curated a diverse internal dataset comprising both dialogues and clean monologue data. The core training corpus consists of 3,000 hours of internal English podcast data, which we processed with 4 steps, as shown in Figure \ref{fig:dataprocessing}. 

First, we applied Microsoft speech enhancement API~\footnote{\url{https://learn.microsoft.com/en-us/azure/ai-services/speech-service/audio-processing-overview}} to remove background noise and music.
Second, we implemented automatic speech recognition and speaker diarization. The diarization results were used to assign speaker identities to each utterance with timestamps. Specifically, we used the Deepgram API~\cite{deepgramEnterpriseVoice}\footnote{\url{https://deepgram.com/}} to obtain word-level transcriptions and speaker labels.
Third, following~\cite{zhang2024covomix}\footnote{\url{https://github.com/vivian556123/NeurIPS2024-CoVoMix/tree/main}}, long dialogues were segmented into shorter clips, each involving two or one speakers, to improve data quality and training efficiency.
Finally, we simulate dialogue segments using the monologue datasets. By pairing utterances from different speakers and introducing controlled overlap or silence ratios, we generate synthetic two-speaker dialogues. Detailed data-processing pipeline is open-sourced~\footnote{\url{https://github.com/vivian556123/covomix2-dataprep.git}}.

For the first training stage, we used the LibriHeavy dataset~\cite{kang2024libriheavy}, comprising 60k hours of high-quality single-speaker audiobook-style recordings. In the second stage, in addition to the podcast dataset, we simulated dialogue-style data by concatenating utterances from different speakers using both LibriHeavy and LibriTTS~\cite{zen2019libritts}. To further enhance the model’s ability to handle overlapping speech, we generated highly overlapped data from LibriHeavy, varying the overlap ratio from 0\% to 100\%.

In order to evaluate the model performance, we design a dialogue test set~\footnote{\url{https://github.com/vivian556123/covomix2-dialogue-testset.git}}, containing 1000 dialogue transcriptions from Dailydialog~\cite{li2017dailydialog} and the acoustic prompts are from Librispeech-test-clean~\cite{panayotov2015librispeech}. We also use samples from this dialogue dataset for subjective evaluation. 

%To ensure that the monologue generation capability is retained, we also designed a monologue test set, with results presented in Appendix~\ref{sec:monologue-perf}.

\subsection{Model Configuration}
In our experiments,  the backbone architecture closely followed the configurations in~\cite{eskimez2024e2}. Specifically, we used Transformer with 24 layers, 16 attention heads, and an embedding dimension of 1024 with U-Net~\cite{unet} style skip connections. The $\sigma_{min}$ is set to $0.1$. We modeled the 100-dimensional log mel-filter bank features, extracted every 10.7 milliseconds from audio samples with a 24kHz sampling rate. A BigVGAN-based~\cite{lee2022bigvgan} vocoder was employed to convert the log mel-filter bank features into waveforms. In addition,  we implemented Classifier-Free Guidance (CFG)~\cite{ho2022classifier} with a dropout probability $p_{uncond} = 20\%$, randomly removing conditioning during training.

In the first training stage, we train the model on 60k hours LibriHeavy~\cite{kang2024libriheavy} dataset for 200k steps with peak learning rate(lr) of 7.5e-5. In the second training stage, we train it for another 200k steps on the combined podcast, audiobook, and simulated dialogue datasets with a peak lr of 5e-5. The model was optimized using the Adam optimizer. A linear-decay learning rate schedule was used in both stages. Each training batch contained two samples, each less than 30 seconds in duration. 

Training was conducted on 32 NVIDIA Tesla V100 GPUs (32GB) with gradient accumulation set to 4. During inference, we used a guidance strength $\alpha$ of 1.0 and performed sampling with 32 function evaluations (NFE) using an ODE solver.

\subsection{Baseline and Evaluation Metrics}
We adopt MoonCast~\cite{ju2025MoonCast}, a latest state-of-the-art dialogue generation model employing a hybrid AR+NAR architecture, as a representative baseline. While simple audio concatenation based on single-speaker models has been proved to be ineffective in capturing speaker interactions~\cite{zhang2024covomix,ju2025MoonCast}, we additionally include Sesame~\cite{sesameCrossingUncanny} as a strong baseline. As a representative model for the CSS task, Sesame generates each utterance sequentially, leveraging previously generated audio as contextual input to maintain coherence across dialogue turns. Detailed baseline configuration comparison is provided in Appendix~\ref{appendix:model}. 

Although other models are capable of dialogue generation, we exclude Dia\cite{githubGitHubNarilabsdia}, CoVoMix\cite{zhang2024covomix}, and NotebookLM~\cite{deepmindNotebookLM} from our comparisons. Dia tends to produce unnaturally rapid and truncated dialogues, and CoVoMix is trained exclusively on 8kHz audio, leading to low-fidelity outputs that are not directly comparable to our high-quality generation setting. NotebookLM is not open-sourced, preventing us from making a reasonable comparison.

To comprehensively assess the recognition accuracy and speaker consistency, we adopt the following objective evaluation metrics: Real-Time Factor (RTF), Word Error Rate (WER), Speaker-Aware Word Error Rate (SA-WER)~\cite{kanda2020joint}, Speaker-Aware Speaker Similarity (SA-SIM) and UTMOS~\cite{baba2024utmosv2}. Specifically, We measure the RTF on a single NVIDIA A100 machine. 
We utilize Microsoft Fast Transcription API~\footnote{\url{https://learn.microsoft.com/en-us/azure/ai-services/speech-service/speech-to-text\#fast-transcription}} as automatic speech recognition and diarization tool to transcribe the generated speech, and we calculate the SA-WER for each word and their corresponding speaker identity. 
The SA-SIM enhances SIM by ensuring that the similarity is computed between embeddings attributed to the speaker identity detected by the diarization model. We utilize WavLM-TDNN~\cite{chen2022wavlm} to extract the speaker embeddings. 

We perform a human evaluation on the generated dialogue examples. We conducted a Comparable Mean Opinion Score (CMOS) experiment to assess user preference in terms of speaker turn handling, interactivity, fluency, and coherence. 15 professional linguistic experts provide  judges for all subjective evaluations. They provide a rating to the second audio, which is randomly selected from a pair of audios, in the (-3 to +3) range. Detailed instructions are in Appendix \ref{appendix:cmos}.

\section{Result and Analysis}
\subsection{Objective and Subjective Evaluation Results }

%\vspace{-0.2cm}

\begin{table}[!ht]
    \centering
    \caption{Model performance comparison on dialogue data}
    \label{tab:objeval}
    \begin{tabular}
    {l|c|cccc|c}
    \toprule
       Model & RTF $\downarrow$  &  WER $\downarrow$& SA-WER $\downarrow$ & SA-SIM $\uparrow$& UTMOS $\uparrow$ & CMOS$\uparrow$ \\ \midrule
         MoonCast & 1.37 & 7.08$\pm$46.23 & 20.40$\pm$51.09 & 0.40$\pm$0.20 &2.65$\pm$0.42 & -0.25$\pm$0.63 \\ 
        Sesame & 2.08 & \textbf{5.62$\pm$5.61} & 9.65$\pm$13.20 & 0.49$\pm$0.17 &2.70$\pm$0.44 & -0.39$\pm$0.40 \\ 
        CoVoMix2 & 0.30 & 5.73$\pm$6.68 & \textbf{6.31$\pm$9.24} & \textbf{0.56$\pm$0.14} &\textbf{3.10$\pm$0.35} & \textbf{0.00$\pm$0.00} \\ \bottomrule
    \end{tabular}
\end{table}

Table~\ref{tab:objeval} presents evaluation results comparing CoVoMix2 against the baseline models MoonCast and Sesame on dialogue test sets. Standard deviations (1-sigma) are reported assuming approximate normality. While WER is positive, some large standard deviations result in negative lower bounds, which are not meaningful in practice but reflect high sample variability. 

Across nearly all metrics, CoVoMix2 demonstrates clear superiority in speech quality, speaker consistency, and inference speed. CoVoMix2 achieves a RTF of 0.30, significantly faster than both MoonCast and Sesame, demonstrating its efficiency as a fully NAR model. 

In terms of content accuracy and speaker consistency, CoVoMix2 attains the best SA-WER and SA-SIM, highlighting its ability to maintain linguistic accuracy and consistent speaker identity across multi-turn interactions.

In contrast, MoonCast, being language-model-based, often suffers from issues such as speaker confusion, hallucinated content, repetitive outputs, and improper termination. These artifacts lead to unstable generation, reflected in its relatively high WER and SA-WER scores. 

Sesame, despite its strong performance in standard WER, occasionally exhibits speaker confusion.  We observe cases where monologue-style outputs include voice characteristics from the other speaker, even though speaker identities are clearly specified. This suggests that the shared contextual input, containing prompts and prior audio from both speakers, may introduce ambiguity, making it difficult for the model to consistently distinguish between speakers in extended dialogue scenarios.

Furthermore, the lower standard deviations observed across all metrics for CoVoMix2 indicate more stable and reliable output quality, further validating the effectiveness of our flow-matching-based architecture for multi-speaker dialogue synthesis.

Finally, the subjective  CMOS comparison between CoVoMix2 and other models demonstrates that our model performs better in terms of speaker turn handling, interactivity, fluency, and coherence with the transcription. We also ask judges to give detailed comments, where better pitch and intonation, better rhythm and less distortion are three main advantages of our proposed CoVoMix2.

\subsection{Speaker Consistency}
%\vspace{-0.2cm}
\begin{figure}[h!]
  \centering
\centerline{\includegraphics[width=1.0\linewidth]{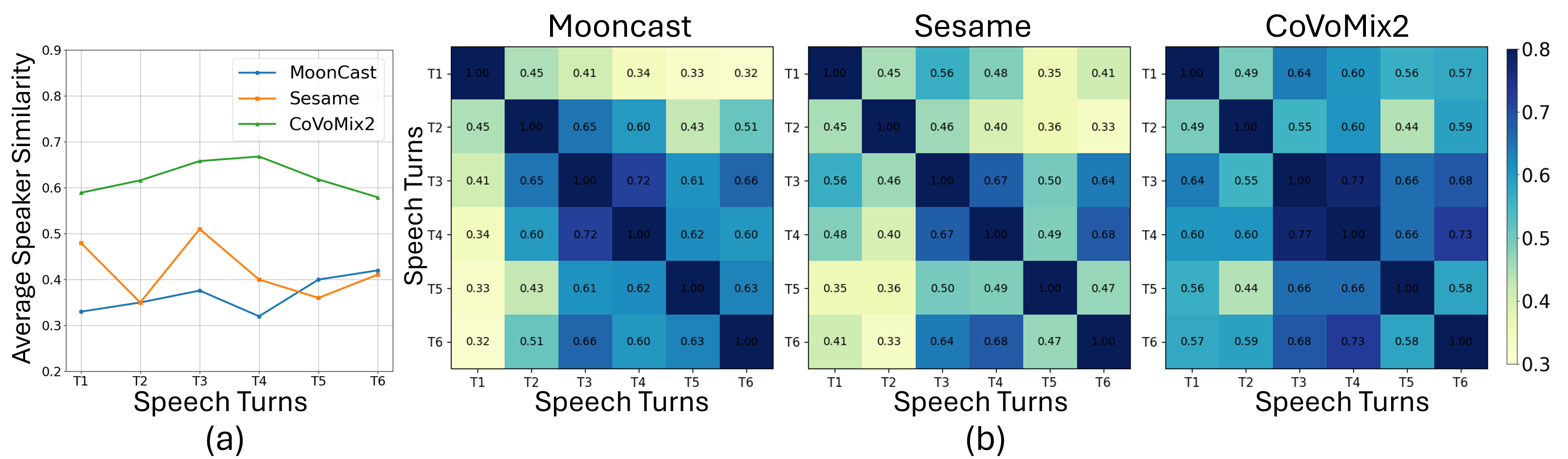}}
\caption{Speaker consistency analysis across dialogue turns.
(a) Average speaker similarity between each generated turn and its corresponding prompt. (b) Pairwise speaker similarity between turns from the same speaker within a dialogue. Consistent color indicates stable speaker timbre across turns. 
}
\label{fig:consistency}
\end{figure}

To evaluate speaker consistency across turns, we selected three long dialogue sequence containing 12 speech turns, with each speaker contributing 6 utterances. 

Figure~\ref{fig:consistency}(a) shows the average speaker similarity between each generated turn and its corresponding prompt. 
While Sesame achieves relatively high similarity in some turns, its performance is inconsistent. This may be due to its design, which generates each utterance independently, relying on a prompt placed only at the beginning. As the dialogue progresses, the prompt information may be forgotten or get confused, especially in longer contexts. 
MoonCast, by contrast, demonstrates consistently low similarity. Although it incorporates prompts at the beginning of each turn, this additional conditioning does not improve speaker accuracy, indicating limited benefit despite the increased computational overhead.

In comparison, CoVoMix2 achieves the highest average similarity with the lowest variance, even though the prompt is provided only once at the beginning of the dialogue. This result demonstrates the robustness in maintaining speaker identity over long conversations.

Figure~\ref{fig:consistency}(b) further investigates intra-dialogue speaker consistency by measuring pairwise speaker similarity across all turns from the same speaker in a long dialogue. A more uniform and consistent color distribution along the rows and columns reflects stronger identity preservation. Both MoonCast and Sesame display noticeable variability, indicating timbre drift or identity shift during generation. In contrast, our model demonstrates consistently high pairwise speaker similarity across all turns, highlighting its effectiveness in preserving speaker timbre throughout multi-turn dialogues.

\subsection{Overlapping Analysis}
\begin{figure}[htbp]
  \centering
\centerline{\includegraphics[width=1.0\linewidth]{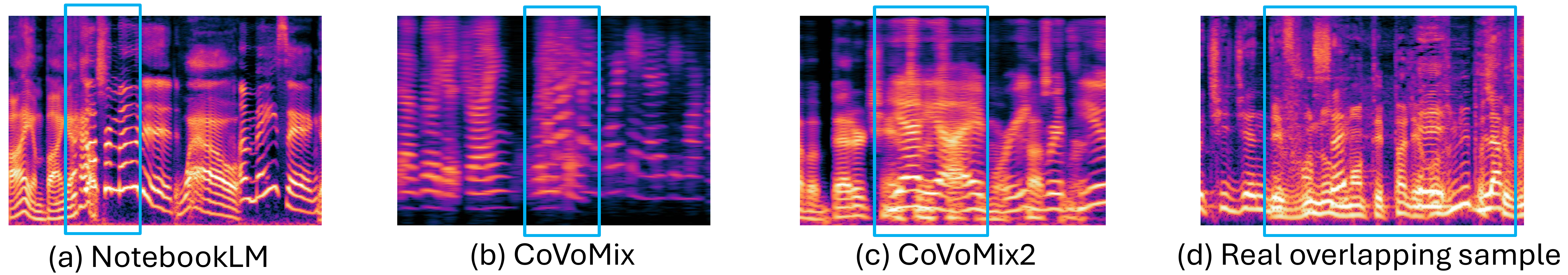}}
\caption{Mel-spectrogram comparison between overlapping samples generated by NotebookLM, CoVoMix, CoVoMix2 and the real sample.
}
\label{fig:overlap-main}
\end{figure}

Achieving overlap in dialogue generation is a challenge for current models. Most models rely on data to achieve overlaps~\cite{zhang2024covomix} and are constrained by the overlap reconstruction capabilities of codecs~\cite{deepmindNotebookLM}.
Figure \ref{fig:overlap-main} shows a visual comparison of mel-spectrograms, extracted from overlapping segments generated by NotebookLM, CoVoMix,   CoVoMix2, and a real overlapping sample, where the first two samples are extracted from the official demo page.  

Overlapping speech is characterized by the simultaneous presence of multiple harmonic structures in the mel-spectrogram, resulting in smooth, continuous spectral patterns. These patterns blend naturally over time, leading to dense, interwoven energy distributions without abrupt boundaries or artificial transitions~\cite{8369155,chen2021overlapped,kristjansson2003high}.
We observe that CoVoMix2 generates overlapping speech with higher spectral fidelity, smoother harmonic structure, and more natural timing alignment, closely matching real overlapping dialogue. In contrast, NotebookLM's output resembles a concatenation of sound segments rather than a genuine learned overlapping process, and 
CoVoMix's speech has low fidelity because of the lower data quality. 

%\vspace{-0.2cm}

\section{Ablation Studies and Extension}

We conducted extensive ablation studies in Appendix~\ref{appendix:ablation} to validate our design choices across data composition, training stages, and special token representation. Results show that pretraining on monologue data is essential for learning stable pronunciation, while data mixing strategy improves generalization to diverse dialogue patterns. Simulated dialogue data may introduce noise. 

Our experiments further demonstrate the robustness of CoVoMix2 to monologue generation, generation with very short prompts, and long dialogue generation. Specifically, it maintains stability even when other baselines fail with prompts under 10 seconds.  Despite being trained on data under 30 seconds, CoVoMix2 is capable of generating stable conversations up to three times longer (approximately 90 seconds). Moreover, we demonstrate that a brief fine-tuning stage using just 20 minutes of human-annotated data can significantly boost transcription accuracy and speaker consistency.

Our framework enables a wide range of practical applications beyond standard dialogue synthesis. Since CoVoMix2 does not require transcriptions for speaker prompts, it supports cross-lingual voice cloning, allowing voices to be transferred across languages. Its temporal control features—including fine-grained manipulation of speech overlap, pauses, and duration—make it especially suited for applications like podcast creation and video dubbing, where precise alignment with visual content or conversational pacing is essential. The ability to generate overlapping speech also enhances realism in multi-speaker scenes, such as dramatic dialogues or animated character interactions.

\section{Conclusion, Limitation, Future Work and Broader Impacts}

In this work, we introduced CoVoMix2, a fully NAR framework for zero-shot multi-talker dialogue generation. By directly predicting mel-spectrograms from disentangled multi-stream transcriptions and leveraging a flow-matching-based method, CoVoMix2 enables efficient, high-quality synthesis of natural, speaker-consistent dialogues, including controlled overlapping speech and fine-grained timing control. Through extensive experiments, we demonstrated that CoVoMix2 outperforms existing models in both speech quality and speaker accuracy while achieving significantly faster inference.

\textbf{Future work} In future work, we plan to extend our framework to support conversations involving more than two speakers, with improved modeling of naturalistic speech overlap and richer conversational dynamics. We also aim to scale up the training to even larger and more diverse datasets, enabling broader generalization across domains and languages.

\textbf{Limitation} While our training data covers a wide range of conversational scenarios, it is primarily automatically transcribed, which may introduce minor inaccuracies, particularly in handling disfluencies such as repetitions or backchannel words. Additionally, since ASR tools lack support for word-level timestamps in overlapping speech, we rely on simulated dialogue data to train overlap scenarios, which can introduce slight deviations in naturalness and degraded audio quality.

\textbf{Broader Impacts} CoVoMix2 offers a versatile and scalable solution for high-quality, human-like speech generation, with potential applications in assistive technology, media production, language learning, and virtual agents. However, since CoVoMix2 could synthesize  speech that maintains speaker identity, it may carry potential risks in misuse of the model, such as spoofing voice identification or impersonating a specific speaker. To mitigate such risks, it is possible to build a detection model to discriminate whether an audio clip was synthesized by CoVoMix2.

% \section*{References}
\newpage
\bibliographystyle{IEEEtran}
\bibliography{mybib}

%%%%%%%%%%%%%%%%%%%%%%%%%%%%%%%%%%%%%%%%%%%%%%%%%%%%%%%%%%%%
\newpage
\appendix
\section{Model Comparison}
\label{appendix:model}
We compare our proposed CoVoMix2 with two baseline models: MoonCast~\cite{ju2025MoonCast}\footnote{\url{https://github.com/jzq2000/MoonCast}} and Sesame~\cite{sesameCrossingUncanny}~\footnote{\url{https://huggingface.co/spaces/sesame/csm-1b}}. The detailed model comparisons are shown in Table \ref{tab:model-detail}, where N/A means not available. 

\begin{table}[!ht]
    \centering
    \caption{Model detail comparison}
    \label{tab:model-detail}
    \setlength\tabcolsep{3pt}
    \begin{tabular}{c|c|ccc|cc|c}
    \toprule
        ~ & ~ &  \multicolumn{3}{c|}{Model Params}  & \multicolumn{2}{c|}{Training Data (hour)} & Training Resource  \\ 
        Model & Type & Backbone & Decoder & Vocoder & Dialogue & Monologue & GPU  \\ \midrule
        MoonCast & AR+NAR & 2.5B & 0.8B & 0.25B & 215k & 300k &  64 A100 80GB  \\ 
        Sesame & AR & 1B & 0.1B & / &\multicolumn{2}{c|}{1000k} & N/A  \\ 
        CoVoMix2 & NAR & 0.3B & / & 0.01B & 3k & 65k & 32 V100 32GB  \\ \bottomrule
    \end{tabular}
\end{table}

\section{Monologue Performance Comparison}
\label{sec:monologue-perf}

To ensure that the dialogue generation models retain strong performance in monologue settings, we evaluate all systems on an audiobook-style monologue test set. The transcription of the audio prompts is obtained using Whisper-large-v3~\cite{radford2022whisper}\footnote{\url{https://huggingface.co/openai/whisper-large-v3}}. As shown in Table~\ref{tab:objevalmono}, our CoVoMix2 trained solely on monologue data, outperforms all baseline models. Notably, after the second-stage training on dialogue data, CoVoMix2 maintains its monologue generation capabilities without much degradation. In contrast, Sesame and MoonCast exhibit degraded performance due to hallucination issues, frequently producing repetitive or semantically irrelevant content. 

In extreme cases, MoonCast fails to terminate the speech generation, resulting in infinite loops. Note that the transcription recognized by Whisper may contain errors, which might be one of the reasons why these language model's hallucination worsens. Our CoVoMix2, however, does not require the transcription corresponding to the prompt, thus effectively avoiding such issues.

\begin{table}[!ht]
    \centering
    \begin{threeparttable}
\captionsetup{justification=centering, width=\linewidth}
    \caption{Model performance comparison on monologue data.}
    \label{tab:objevalmono}
    \begin{tabular}{c|ccc}
    \toprule
       Model  & WER $\downarrow$ & SIM$\uparrow$ & UTMOS $\uparrow$\\ \midrule
         MoonCast & 15.92$\pm$28.50 & 0.63$\pm$0.16 & 2.82$\pm$0.48  \\ 
        Sesame  & 7.58$\pm$12.39 & 0.72$\pm$0.17 & 2.81$\pm$0.52 \\ 
        CoVoMix2$\dagger$ &  \textbf{4.30$\pm$4.75} & \textbf{0.78$\pm$0.09} & 2.97$\pm$0.35  \\
        \midrule
        CoVoMix2 & 4.45$\pm$4.19& 0.66$\pm$0.17 & \textbf{3.19$\pm$0.37}  \\ \bottomrule
    \end{tabular}
    \begin{tablenotes}
        \footnotesize
        \item[$\dagger$] CoVoMix2$\dagger$ is only pre-trained on the monologue data.
    \end{tablenotes}
    \end{threeparttable}
\end{table}

\section{Extended Dialogue Performance Comparison}
\subsection{Dialogue Performance Comparison with Prompts of different length}

\begin{figure}[htbp]
  \centering
\centerline{\includegraphics[width=1.0\linewidth]{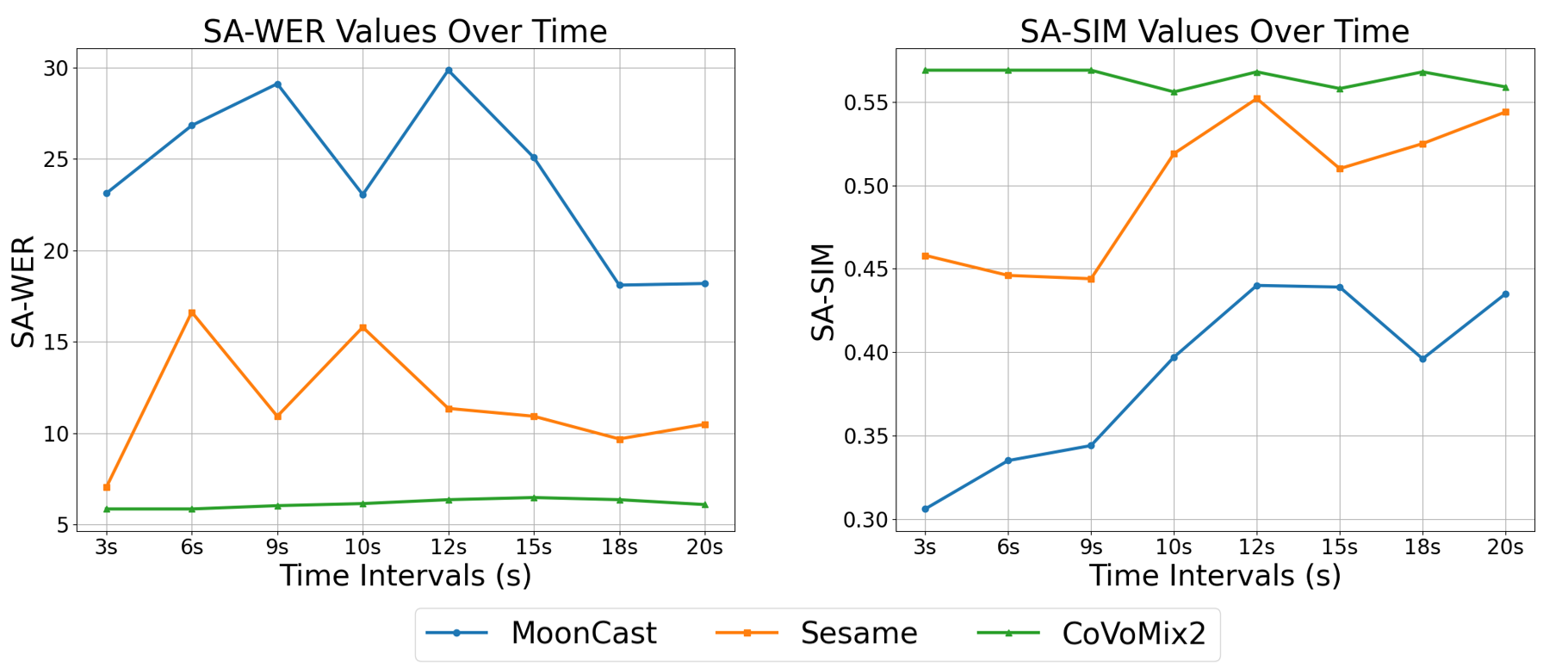}}
\caption{SA-WER and SA-SIM comparison with prompt of different lengths}
\label{fig:prompt-length}
\end{figure}

To demonstrate that our model maintains a stable advantage even with very long or very short prompts, we conducted an additional experiment. We selected 20 speech clips, each longer than 20 seconds, from 20 different speakers in the LibriSpeech dataset~\cite{panayotov2015librispeech}. We then trimmed these clips to lengths of 3 to 18 seconds. For transcription, we used Whisper-large-v3, and for the text component, we used 100 dialogues from the Dailydialog dataset. As shown in Figure \ref{fig:prompt-length}, our model's SA-WER and SA-SIM performance remains consistently strong across all prompt lengths. This further confirms that our model performs reliably well, regardless of how long the input prompt is.

\subsection{Dialogue Performance Comparison with Different Synthesized Length}

Due to resource constraints, our training data currently consists of segments less than 30 seconds in duration. Nevertheless, our model demonstrates the ability to infer dialogues longer than 30 seconds, and our test set includes numerous dialogues exceeding one minute. 

To further investigate long dialogue generation, we manually designed four dialogues with varying durations. As shown in Table \ref{table:length_compare}, we observed that both CoVoMix2 and Mooncast exhibit degraded performance after 90 seconds, while Sesame fails to generate dialogues longer than 120 seconds. Additionally, Sesame showed unstable results even for dialogues under 30 seconds, primarily due to speaker confusion issues.

\begin{table}[h!]
\centering
\caption{Comparison of SA-WER/SA-SIM with Different Synthesized Length}
\label{table:length_compare}
\begin{tabular}{c|c|c|c|c}
\toprule
System & 30s & 60s & 90s & 120s \\ 
 \midrule
Mooncast& 5.81/0.647 & 4.57/0.576 & 13.95/0.487 & 19.75/0.516 \\
Sesame & 24.41/0.677 & 5.22/0.585 & 3.98/0.263 & Failed \\
CoVoMix2 & 6.97/0.686 & 5.88/0.648 & 14.74/0.618 & 14.64/0.627 \\ 

\bottomrule
\end{tabular}
\end{table}

\subsection{Dialogue Performance Comparison under Real-world Scenarios}
We designed an additional test set to use more realistic conversation style speeches as prompts. The speaker prompts for this new set are derived from 10 distinct real-life dialogues from the NCSSD dataset-CEN~\cite{liu2024generative}. This dataset is collected from the internet and features prompts with natural conversational prosody, including various acoustic scenarios such as background noise. The corresponding text content for this test set is sourced from the DailyDialog dataset~\cite{li2017dailydialog}.

As presented in Table \ref{table:real}, our model, CoVoMix2, consistently demonstrates superior performance compared to baseline models even under these challenging real-world conditions.

\begin{table}[h!]
\centering
\caption{Performance Comparison under Real-world Scenarios}
\label{table:real}
\begin{tabular}{c|c|c|c}
\toprule
System & SA-WER & SA-SIM & UTMOS \\ 
 \midrule
Mooncast& 26.26 &	0.276 &	2.09 \\
Sesame & 25.19 &	0.250 &	1.80 \\
CoVoMix2 & 9.32 &	0.322 &	2.89
 \\ 
\bottomrule
\end{tabular}
\end{table}

\section{Ablation Studies}
\label{appendix:ablation}
To better understand the design choices and training strategies that contribute to the performance of CoVoMix2, we conducted a series of ablation studies covering data mixing, training stages, and the handling of silence tokens. 

\subsection{Data Mixing Strategy}

We examine how various combinations of training data affect performance. Table~\ref{tab:data} presents results on both monologue and dialogue test sets, comparing models trained on different subsets of the data. We observe that in addition to dialogue podcast data, incorporating monologue podcast data significantly improves performance. Adding LibriTTS and LibriHeavy audiobook datasets provides further gains, with LibriTTS showing slightly better results due to its cleaner and more accurately aligned transcriptions. Interestingly, simulated dialogue data (created by combining monologues with a certail ratio of overlap or silence) does not consistently improve performance and may even slightly degrade it. This is likely due to distribution mismatch or artifacts introduced by the simulation process. However, simulated dialogues are still necessary to enable overlapping speech generation capability, as ASR-transcribed real-world data rarely provides accurate overlapping timestamps.

\begin{table}[!ht]
    \centering
        \caption{Ablation Study of data}
    \label{tab:data}
\setlength\tabcolsep{2.5pt}
    \begin{tabular}{c|ccccc|ccc|cccc}
   \toprule
        ~ & \multicolumn{5}{c|}{Training Data}  & \multicolumn{3}{c|}{Monologue} & \multicolumn{4}{c}{Dialogue}    \\
        ID & Dia & Mono & Simu & LH & LT & WER$\downarrow$ & SIM$\uparrow$ & UTMOS$\uparrow$ & WER$\downarrow$ & SA-WER$\downarrow$ & SA-SIM$\uparrow$ & UTMOS$\uparrow$  \\ \midrule
  1 &\ding{51} &\ding{55} &\ding{55} &\ding{55} &\ding{55} & 8.58 & 0.65 & 3.12 & 8.11 & 8.63 & 0.50 & 2.97  \\ 
        2 &\ding{51} &\ding{51} &\ding{55} &\ding{55} &\ding{55} & 6.30 & 0.61 & 3.18 & 7.89 & 8.51 & 0.50 &   3.00\\ 
        3 &\ding{51} &\ding{51} &\ding{51} &\ding{55} &\ding{55} & 7.69 & 0.64 & 3.10 & 8.72 & 9.31 & 0.51 &   2.99\\ 
        4 &\ding{51} &\ding{51} &\ding{51} &\ding{51} &\ding{55} & 6.18 & 0.66 & 3.24 & 7.08 & 7.43 & 0.55 &   3.06\\ 
        5 &\ding{51} &\ding{51} &\ding{51} &\ding{55} &\ding{51} & 5.37 & 0.65 & 3.21 & 6.00 & 6.58 & 0.57 &   3.10\\ 
        6 &\ding{51} &\ding{51} &\ding{55} &\ding{55} &\ding{51} & 5.42 & 0.65 & 3.27 & 5.81 & 6.27 & 0.56 &   3.13\\ 
 \bottomrule
    \end{tabular}
\end{table}

\subsection{Training Stages}
\label{sec:finetune}
CoVoMix2 is trained using a two-stage training pipeline, with an optional third fine-tuning stage that can further improve performance. To assess the benefit of the first and the third stage, we conduct an ablation study by eliminating the first stage and fine-tuning the model for 2,000 steps on just 20 minutes of human-annotated dialogue data. Although this fine-tuning data is small in scale, it is highly accurate and clean, in contrast to the ASR-transcribed training data.

Table~\ref{tab:training-stage} shows results on a test set featuring the same two speakers as in the fine-tuning data. Models trained without the pretraining perform very poorly, with WER and SA-WER exceeding 80\%, highlighting the critical importance of the first stage.  Fine-tuning with just 2,000 steps on a small amount of human-annotated dialogue data significantly improves accuracy, especially in WER and SA-WER. This suggests that even limited high-quality data can help correct transcription noise learned from ASR-transcribed training sets.
 
\begin{table}[!ht]
    \centering
    \caption{Impact of training stages on a two-speaker test set}
    \label{tab:training-stage}
    \begin{tabular}{c|c|cccc}
\toprule
        Pre-train & Fine-tune & WER$\downarrow$ & SA-WER$\downarrow$ & SA-SIM$\uparrow$ & UTMOS$\uparrow$  \\ \midrule
        \ding{55}  & \ding{55} & 82.66  & 92.40  & 0.29  & 3.00   \\ 
        \ding{51}  & \ding{55} & 5.67  & 6.21 & 0.53  & 3.48   \\ 
        \ding{51}  & \ding{51}  & 3.66 & 3.81 & 0.56  & 3.35  \\  \bottomrule
    \end{tabular}
\end{table}

\subsection{Silence Token Representation}

Our input text streams include not only characters but also two types of special tokens: a continuation token $[P]$ and a silence token $[S]$. We explore three options. 1) using $[P]$ for both continuation and silence. 2) using a generic $[S]$ for silence. 3) using speaker-aware silence tokens $[S1]$ and $[S2]$. Table~\ref{tab:silencetoken} summarizes the results. We find that using a separate silence token $[S]$ improves dialogue accuracy and controllability over using $[P]$ alone. However, using speaker-aware silence tokens does not provide significant benefits, and in some cases slightly degrades dialogue performance, possibly due to over-specification.

\begin{table}[!ht]
    \centering
     \caption{Impact of different silence token representation}
    \label{tab:silencetoken}
    \begin{tabular}{c|ccc|cccc}
    \toprule
    ~ & \multicolumn{3}{c|}{Monologue}  & \multicolumn{4}{c}{Dialogue}  \\
        Silence Token  & WER$\downarrow$  & SIM$\uparrow$ & UTMOS$\uparrow$ & WER$\downarrow$ & SA-WER$\downarrow$ & SA-SIM$\uparrow$ & UTMOS$\uparrow$  \\ \midrule
     $[P]$ & 5.26 &0.64&	3.15&6.45&7.09&0.55&3.08 \\
     $[S]$&6.83&0.65&3.14&5.22&5.59&0.56&3.02\\
     $[S1,S2]$ & 5.37&0.65&3.21&6.00&6.58&0.57&3.10 \\\bottomrule
    \end{tabular}
\end{table}

\section{Subjective Evaluation}
\label{appendix:cmos}

Table \ref{tab:cmos-instruction} shows the Comparative Mean Opinion Score (CMOS) Evaluation instruction. 15 professional linguistic experts provide
 judges for this CMOS evaluation. They provide a rating of preference in the (-3 to +3) range, given two audios with the same transcription and the speaker prompts.

\begin{table}[h!]
\centering
\caption{Comparative Subjective Mean Opinion Score (CMOS) Evaluation Instructions}
\label{tab:cmos-instruction}
\begin{tabular}{p{3cm} p{10cm}} % Split into two columns
\toprule
\multicolumn{2}{p{13cm}}{\textbf{Instruction}} \\ \midrule
\multicolumn{2}{p{13cm}}{This is to compare two AI podcast audio. Listen to both audios as you are listening to a real human podcast and give your preference considering the following aspects.} \\ 
\multicolumn{2}{p{13cm}}{1. Speaker Attribution Accuracy : Is each utterance spoken by the correct speaker as indicated in the transcription? Does the voice match the intended identity?} \\
\multicolumn{2}{p{13cm}}{2. Speaker Turn Handling: Are speaker changes handled correctly and clearly? Does the transition between speakers align with the dialogue flow?} \\
\multicolumn{2}{p{13cm}}{3. Speaker Consistency:  Does each speaker maintain a consistent voice, tone, and speaking style throughout the clip?} \\
\multicolumn{2}{p{13cm}}{4. Interactivity and Fluency: Does the conversation sound natural and interactive? Are the responses well-timed and appropriate, without awkward pauses or overlaps?} \\
\multicolumn{2}{p{13cm}}{5. Coherence with the Transcription: Does the spoken content accurately follow the given transcription, including emotion, prosody, and speaking style?} \\ \midrule
\multicolumn{2}{p{13cm}}{\textbf{Which one is better?}} \\
\makecell[r]{-3:} & Dialogue 1 is much better \\
\makecell[r]{-2:} & Dialogue 1 is better \\
\makecell[r]{-1:} & Dialogue 1 is slightly better \\ 
\makecell[r]{0:} & Can't tell which is better \\
\makecell[r]{1:} & Dialogue 2 is slightly better \\
 \makecell[r]{2:} & Dialogue 2 is better \\
\makecell[r]{3:} & Dialogue 2 is much better \\
\bottomrule
\end{tabular}
\end{table}

\end{document}